\documentclass[aps,twocolumn,showkeys,nofootinbib]{revtex4-1}

\usepackage{graphicx,amsmath,amssymb,amsfonts,latexsym,color,dcolumn,bm,mathtools,dsfont}
\usepackage{verbatim}
\usepackage{epstopdf}

\providecommand{\abs}[1]{\left\lvert#1\right\rvert}

\providecommand{\ie}[0]{\textit{i.e.~}}
\providecommand{\eg}[0]{\textit{e.g.~}}

\definecolor{blue}{rgb}{0.223,0.223,0.667}
\definecolor{red}{rgb}{0.7,0,0}

\begin{document}


\title{Microwave measurement beyond the quantum limit with a nonreciprocal amplifier}

\author{F. Lecocq$^{1}$, L. Ranzani$^2$, G. A. Peterson$^{1}$, K. Cicak$^1$, A. 
Metelmann$^3$, S. Kotler$^{1}$, R. W. Simmonds$^1$, J. D. Teufel$^1$ and J. 
Aumentado$^1$}
\affiliation{$^1$National Institute of Standards and Technology, 325 Broadway, 
	Boulder, CO 80305, USA}
\affiliation{$^2$Raytheon BBN Technologies, Cambridge, Massachusetts 02138, USA}
\affiliation{$^3$Dahlem Center for Complex Quantum Systems and Fachbereich Physik, Freie Universitat Berlin, 14195 Berlin, Germany}
\date{\today}

\begin{abstract}
The measurement of a quantum system is often performed by encoding its state in a single observable of a light field. The measurement efficiency of this observable can be reduced by loss or excess noise on the way to the detector. Even a \textit{quantum-limited} detector that simultaneously measures a second non-commuting observable would double the output noise, therefore limiting the efficiency to $50\%$. At microwave frequencies, an ideal measurement efficiency can be achieved by noiselessly amplifying the information-carrying quadrature of the light field, but this has remained an experimental challenge. Indeed, while state-of-the-art Josephson-junction based parametric amplifiers can perform an ideal single-quadrature measurement, they require lossy ferrite circulators in the signal path, drastically decreasing the overall efficiency. In this paper, we present a nonreciprocal parametric amplifier that combines single-quadrature measurement and directionality without the use of strong external magnetic fields. We extract a measurement efficiency of $62_{-9}^{+17} \%$ that exceeds the quantum limit and that is not limited by fundamental factors. The amplifier can be readily integrated with superconducting devices, creating a path for ideal measurements of quantum bits and mechanical oscillators.
\end{abstract}

\pacs{Valid PACS appear here}
\maketitle

  
\section{Introduction}

To measure the state of a quantum system, one typically entangles it with an auxiliary quantum system, a `meter', that in turn can be easily measured with a classical apparatus \cite{Haroche2006,IntroQnoise2010}. Non-idealities in the meter measurement, such as dissipation and excess noise, are quantified by a single parameter: the measurement efficiency, $0\leq\eta\leq1$. While state estimation can be achieved with high fidelity even with limited measurement efficiency \cite{Harty2014HighFidelityIon}, a highly efficient measurement is critical for controlling the quantum system, enabling, for example, analog quantum feedback \cite{Vijay2012,deLange2014,Rossi2018} and measurement-based entanglement \cite{Roch2014,Liu2016}.

In superconducting microwave devices, the quantum information  --- \eg the state of a quantum bit or the position of a mechanical oscillator --- is typically encoded in a single quadrature of a microwave field \cite{Blais2007,OptoReview2014}. The measurement of this meter takes the form of amplification, necessary to overcome the noise of room temperature electronics. State-of-the-art measurement schemes rely on Josephson-junction based parametric amplifiers as the first stage of amplification \cite{teufel2009nanomechanical,Kindel2016,Hatridge2013,Liu2016,Vijay2012,walter2017rapid}. While these amplifiers can \textit{in principle} have ideal noise performance, in practice they require additional components in the signal path that significantly reduce the overall system measurement efficiency. Most importantly, these amplifiers are reciprocal and therefore require multiple microwave circulators to route amplified microwave signals away from the quantum system and toward the measurement apparatus. Microwave circulators break reciprocity using the Faraday effect induced by large magnetic fields\cite{Fay1965} that are incompatible with superconducting devices, thus requiring additional wiring with long coaxial cables and multiple connectors to physically separate them from both the quantum system and the amplifier. Highly optimized measurement chains have demonstrated efficiencies around $0.5$ to $0.7$ \cite{Kindel2016,walter2017rapid}, leaving little room for further improvement.

Recently, amplifiers breaking reciprocity without the use of strong magnetic fields have been developed, in an effort to directly integrate them with superconducting quantum devices to maximize measurement efficiency. Among these efforts are traveling wave amplifiers \cite{macklin2015near} and multi-mode amplifiers \cite{abdo2013,sliwa2015reconfigurable,lecocq2017nonreciprocal,Chapman2017Circulator,Lepinay2019}. While all of these approaches meet the requirements for gain, bandwidth, dynamic range, and integrability, they also amplify, \textit{i.e.} measure, both non-commuting quadratures of the microwave field (known as phase-preserving amplification). As a result, the output noise doubles \cite{Caves1980}, effectively putting a quantum limit on the measurement efficiency at $0.5$ \cite{Korotkov2016}. Realizing a single quadrature measurement (known as phase-sensitive amplification) with a nonreciprocal amplifier has been a long-standing challenge and has, so far, only been proposed theoretically \cite{metelmann2015nonreciprocal}.

Here we report on the novel theory and first experimental implementation of a phase-sensitive nonreciprocal amplifier. Based on a Field-Programmable Josephson Amplifier (FPJA)\cite{lecocq2017nonreciprocal}, it utilizes four parametric pumps to achieve $\approx24~\text{dB}$ of quadrature gain with a $7~\text{MHz}$ bandwidth, near-unity transmission in the reverse direction, with both input and output impedance matched to the $50~\Omega$ environment. Using a self-calibrated noise source we extract a measurement efficiency of $\eta= 0.62_{-0.09}^{+0.17}$, exceeding the quantum limit of $0.5$ for phase-preserving amplification. Straightforward improvements could enable efficiency of more that $0.95$.

\begin{figure*}
	\includegraphics[scale=1]{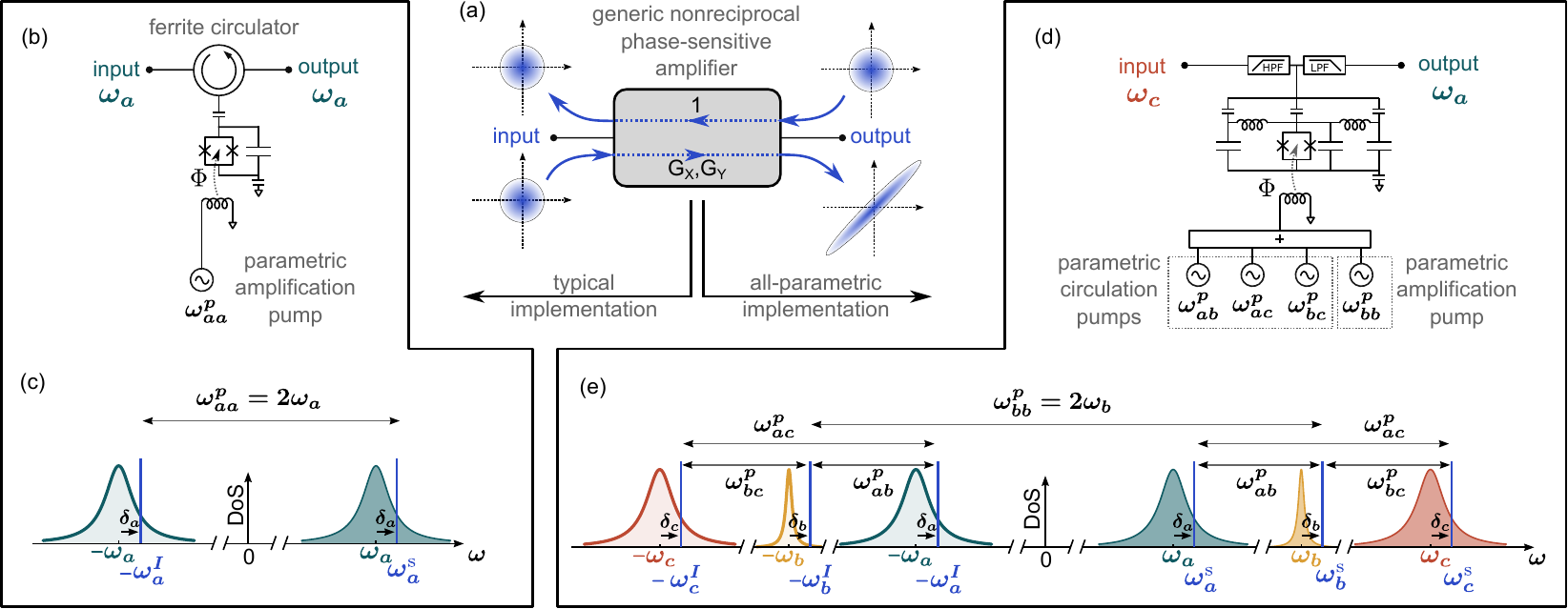}
	\caption{(a) Schematic of a generic nonreciprocal phase-sensitive amplifier. A signal incident at the input is phase-sensitively amplified toward the output, while in the reverse direction, a signal is transmitted from output to input with unity transmission. (b) Typical physical implementation of (a) using a Josephson Parametric Amplifier (JPA), providing phase-sensitive reflection gain, and a ferrite circulator, ensuring  signal flow from input to output. (c) Frequency-space diagram showing the density of states (DoS) for the JPA. A strong pump of frequency $\omega_{aa}^p = 2\omega_{a}$ amplifies an input signal at the frequency $\omega_{a}^\text{S} = \omega_{a}+\delta_a$ and generates an idler at the frequency $-\omega_{a}^\text{I} = -\omega_{a}+\delta_a$. (d) All-parametric physical implementation of (a). The ferrite circulator in (b) is replaced by a parametric circulator based on a Field Programmable Josephson Amplifier (FPJA). This device has three resonant modes: an `input' mode of frequency $\omega_{c}$, a `gain' mode of frequency $\omega_{b}$ and an `output' mode of frequency $\omega_{a}$. Input and output signals are routed to different physical ports by low-pass and high-pass filters (LPF and HPF, respectively). (e) Frequency-space diagram showing the density of states for the FPJA. Similarly to (c), a strong pump of frequency $\omega_{bb}^p = 2\omega_{b}$ amplifies a signal and idler at the frequency $\omega_{b}^\text{S} = \omega_{b}+\delta_b$ and $-\omega_{b}^\text{I} = -\omega_{b}+\delta_b$. Parametric circulation is obtained using three frequency conversion pumps with frequencies $\omega_{ab}^p = \omega_{b}-\omega_{a}$, $\omega_{bc}^p = \omega_{c}-\omega_{b}$, $\omega_{ac}^p = \omega_{c}-\omega_{a}$, satisfying the condition  $\omega_{ab}^p + \omega_{bc}^p = \omega_{ac}^p$ and connecting signals and idlers at $\omega_{j}^\text{S} = \omega_{j}+\delta_j$ and $-\omega_{j}^\text{I} = -\omega_{j}+\delta_j$ (${j} \in \{a,b,c\}$).
	\label{fig1}}
\end{figure*}

\section{Concept}

Conceptually, the device in this work can be understood by analogy with the typical combination of a Josephson Parametric Amplifier (JPA) and a traditional 3-port circulator, see Fig.~\ref{fig1}. Both systems implement the same generic scattering parameters: a signal incident at the input gets phase-sensitively amplified toward the output, whereas a signal incident at the output is simply routed with unity transmission to the input, see Fig.~\ref{fig1}(a).

A JPA consists of a single resonant circuit whose frequency $\omega_{a}$ is tunable by the flux applied to a SQUID, Fig.~\ref{fig1}(b) and (c). A strong pump modulates the flux at a frequency $\omega_{aa}^p\approx2\omega_{a}$, mediating the mixing process between a signal and an idler of frequency $\omega_{a}^\text{S}$ and $\omega_{a}^\text{I}$, with conservation of energy dictating $\omega_{aa}^p=\omega_{a}^\text{S}+\omega_{a}^\text{I}$. Physically, pump photons enable the creation and annihilation of correlated pairs of signal and idler photons, leading to direct and idler gain \cite{yurke1989observation,castellanos2007widely}. Importantly, as the signal and idler tones exist within the same resonance of the system, it is natural to consider their linear combinations, \ie quadratures of the resonator's field. The correlations between signal and idler inherent to the mixing process lead to quadrature-sensitive gain \cite{yurke1989observation,castellanos2007widely}. However, this amplifier is reciprocal and therefore requires a microwave circulator to control the signal flow. In a typical microwave circulator, three ports are coupled by a circular ferrite resonator biased by a large magnetic field. Input signals acquire a phase that, due to the Faraday effect, depends on whether they propagate clockwise or counterclockwise, leading to constructive interference at the output port and destructive interference at the isolated port \cite{Fay1965,Pozar2012}.

By contrast, here we engineer this behavior by programming a single FPJA with four parametric pumps to realize both circulation and phase sensitive amplification, without a strong magnetic field, see Fig.~\ref{fig1}(d) and (e). The FPJA is a circuit with three resonances (`modes') whose frequencies $\omega_{a}$, $\omega_{b}$, and $\omega_{c}$ are tunable by the flux applied to a single SQUID. As in a JPA, a strong pump of frequency $\omega_{bb}^p\approx2\omega_{b}$ couples a signal at $\omega_{b}^\text{S}$ to its idler at $\omega_{b}^\text{I}$, inducing quadrature-sensitive gain. Circulation is ensured by three pumps, of frequencies $\omega_{ab}^p \approx \omega_{b}-\omega_{a}$, $\omega_{bc}^p \approx \omega_{c}-\omega_{b}$ and $\omega_{ac}^p \approx \omega_{c}-\omega_{a}$. These pumps provide the energy needed for the coherent exchange of signal photons between all three modes (frequencies $\omega_{j}^\text{S}$, with ${j} \in \{{a,b,c}\}$), as well as for the coherent exchange of idler photons between all three modes (frequencies $\omega_{j}^\text{I}$, with ${j} \in \{{a,b,c}\}$). Importantly these pumps must satisfy the condition $\omega_{ab}^p + \omega_{bc}^p = \omega_{ac}^p$, forming a loop in frequency space. As in a ferrite circulator, signals traveling clockwise or counter clockwise acquire a different phases, now set by the pump phases $\phi_\mathrm{loop}=\phi_{ab}+\phi_{bc}-\phi_{ac}$ \cite{lecocq2017nonreciprocal}. As an example, for $\phi_\mathrm{loop}=\pi/2$, an input signal near mode $c$ is converted toward mode $b$, amplified, and converted to the output mode $a$. In the reverse direction an input signal near mode $a$ is scattered to mode $c$ without gain. Finally, input and output signals are routed to different physical ports using on-chip frequency filters. A complete theoretical description and details about device fabrication and amplifier tuning can be found in appendix.

\section{Scattering parameters}

To calculate the scattering parameters of the device we solve the coupled equations of motion (EoM) for the field amplitudes in the Fourier domain, following the graph-based method discussed in \cite{ranzani2015graph,lecocq2017nonreciprocal}. We define the vector of intra-cavity field amplitudes $\mathbf{A}=(a^\text{S},b^\text{S},c^\text{S},a^\text{I*},b^\text{I*},c^\text{I*})^\top$, input field amplitudes $\mathbf{A}_{\text{in}}=(a_\text{in}^\text{S},b_\text{in}^\text{S},c_\text{in}^\text{S},a_\text{in}^\text{I*},b_\text{in}^\text{I*},c_\text{in}^\text{I*})^\top$, ouput field amplitudes $\mathbf{A}_{\text{out}}=(a_\text{out}^\text{S},b_\text{out}^\text{S},c_\text{out}^\text{S},a_\text{out}^\text{I*},b_\text{out}^\text{I*},c_\text{out}^\text{I*})^\top$, diagonal matrices for the total loss rates $\mathbf{K}=\text{diag}(\sqrt{\kappa_a},\sqrt{\kappa_b},\sqrt{\kappa_c},\sqrt{\kappa_a},\sqrt{\kappa_b},\sqrt{\kappa_c})$, external couplings $\mathbf{K}^{\text{ext}}=\text{diag}(\sqrt{\kappa_a^{\text{ext}}},\sqrt{\kappa_b^{\text{ext}}},\sqrt{\kappa_c^{\text{ext}}},\sqrt{\kappa_a^{\text{ext}}},\sqrt{\kappa_b^{\text{ext}}},\sqrt{\kappa_c^{\text{ext}}})$ and coupling efficiencies $\mathbf{H}=\text{diag}(\sqrt{\eta_a},\sqrt{\eta_b},\sqrt{\eta_c},\sqrt{\eta_a},\sqrt{\eta_b},\sqrt{\eta_c})$, where  $\eta_{j}=\kappa_{j}^{\text{ext}}/\kappa_{j}$. The EoM can be expressed as $\mathbf{K}\mathbf{M}\mathbf{K}\mathbf{A} = i\mathbf{K}^{\text{ext}}\mathbf{A}_{\text{in}}$, where  $\mathbf{M}$ is the mode-coupling matrix:

\begin{equation}
\mathbf{M}=
\left(\begin{array}{ccc|ccc}
\ \Delta_{a}^S   & \   \  \beta_{ab}  &\ \ \beta_{ac}   &     0           &        0        &      0     \\
\ \beta_{ab}^{*} & \ \Delta_{b}^S   &\ \  \beta_{bc}   &     0           &    \beta_{bb}   &      0     \\
\ \beta_{ac}^*   & \ \beta_{bc}^{*} &\ \  \Delta_{c}^S &     0           &        0        &      0     \\ \hline
\ 0         &  \  \     0        &\   \     0       & -{\Delta_a^I}^* & -\beta_{ab}^*   & -\beta_{ac}^* \\
\ 0         & \ -\beta_{bb}^*  &\   \     0       & -\beta_{ab}     & -{\Delta_b^I}^* & -\beta_{bc}^* \\
\ 0         &  \  \     0        &  \  \    0       & -\beta_{ac}     & -\beta_{bc}     & -{\Delta_c^I}^*
\end{array}\right).
\label{eq:BTmatrix}
\end{equation}

The diagonal elements are the normalized complex detuning terms $\Delta_j^{S,I}=(\omega_{j}^\text{S,I}-\omega_{j})/\kappa_j+i/2$ and the off-diagonal elements are the normalized coupling terms 
$\beta_{jk}=g_{jk}/(2\sqrt{\kappa_j\kappa_k})$ between modes ${j}$ and ${k}$, with ${j},{k} \in \{{a,b,c}\}$. Each $g_{jk}=\abs{g_{jk}}e^{i\phi_{jk}}$ is the parametrically induced coupling rate, proportional to the amplitudes of the pump with frequency $\omega_{jk}^p$ and phase $\phi_{jk}$. We emphasize the block structure of $\mathbf{M}$, which reflects the coupling networks and frequency space diagram in Fig.~\ref{fig1}(d): the diagonal blocks correspond to circulation between the signals and between the idlers, and the anti-diagonal blocks correspond to amplification between signal and idler in mode $b$. Solving these EoM yields the scattering matrix $\mathbf{S}$, defined as $\mathbf{A}_{\text{out}}=\mathbf{S}\mathbf{A}_{\text{in}}$:
\begin{equation}
\mathbf{S}= i\mathbf{H}\mathbf{M}^{-1}\mathbf{H}-\mathds{1}.
\label{eq:Sdef}
\end{equation}

The device behavior is programmed in four steps. First, each frequency conversion pump is individually calibrated to match the conversion rates to the  dissipation rates $\abs{g_{jk}}=\sqrt{\kappa_j\kappa_k}$, \ie $\beta_{jk}=1/2$. Second, all three frequency conversion pumps are turned on, forming an interferometer that completely cancels propagation in the clockwise or counter-clockwise direction depending on sum of the pump phases  $\phi_\mathrm{loop}=\phi_{ab}+\phi_{bc}-\phi_{ac}=\pm\pi/2$ \cite{lecocq2017nonreciprocal}. Third, the frequency conversion rates to mode $b$, $\beta_{ab}$ and $\beta_{bc}$, are increased in order to overwhelm the loss rate of mode $b$, while maintaining $|\beta_{ab}|=|\beta_{bc}|$. This yields high reflection off of mode $b$ and the scattering parameters resemble the case of a JPA on a circulator in the absence of gain. As we will discuss later, this step is critical to obtain low added noise, preventing the signal of interest from dissipating in mode $b$ and being replaced by uncorrelated vacuum noise. Finally, the amplification pump is turned on. In the ideal resonant case, neglecting internal loss and for $\phi_\mathrm{loop}=\pi/2$, the simplified scattering matrix $\mathbf{S_{ac}}$ between field amplitudes $a^\text{S}$ and $c^\text{S}$ is:
\begin{equation}
\mathbf{S_{ac}}=
\begin{pmatrix}
    0         &    \sqrt{G_\text{S}}   \\
    1         &      0     
\end{pmatrix},
\label{eq:Sac}
\end{equation}
where $\sqrt{G_\text{S}}=(2s+r^{2}-1)/(1-r^{2})$ and 
\begin{align}
s & =\frac{4|\beta_{ab}|^{2}}{1+4|\beta_{ab}|^{2}},\\
r & =\frac{2|\beta_{bb}|}{1+4|\beta_{ab}|^{2}},
\end{align}
with $0\leq r,s\leq1$. Physically, $s$ represents the ratio between the conversion rates to mode $b$ and the total dissipation rate of mode $b$, while $r$ represents the ratio between the amplification rate of mode $b$ and the total dissipation rate of mode $b$. 

\begin{figure}
	\includegraphics[scale=1]{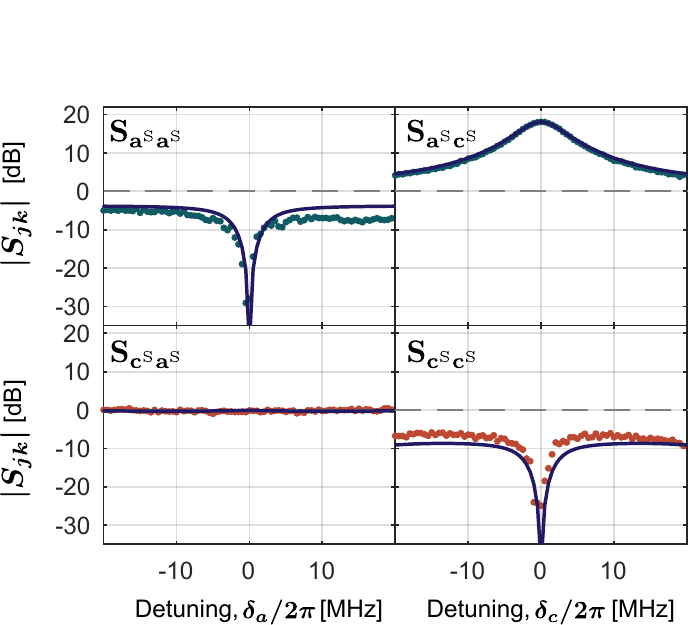}
	\caption{Measured scattering parameters of the FPJA programmed as a phase-sensitive directional amplifier (the dots) as function of the drive detuning $\delta_{a,c}=(\omega_{a,c}^\text{S}-\omega_{a,c})$. The solid lines are the predictions from Eq.\ref{eq:Sdef}. The device exhibits high gain from mode $c$ to mode $a$, over a $7~\text{MHz}$ bandwidth, and unity transmission in the reverse direction. The device is impedance matched, with return loss exceeding  $10~\text{dB}$ over a $4~\text{MHz}$ bandwidth.
		\label{fig2}}
\end{figure}

Experimentally we start by fixing a flux bias point and measure the modes frequencies $\left(\omega_{a},\omega_{b},\omega_{c}\right)/2\pi= \left(6.876,7.932,10.782\right)~\text{GHz}$, linewidths $\left(\kappa_{a},\kappa_{b},\kappa_{c}\right)/2\pi= \left(83,15,45\right)~\text{MHz}$ and coupling parameters $\eta_a=0.99$ and $\eta_c=0.99$. We then proceed to the four programming steps described above, by comparing the measured scattering parameters to the solutions of Eq.\ref{eq:Sdef}. The results are shown in Fig.~\ref{fig2}, for $|\beta_{ac}|=0.5$, $|\beta_{ab}|=|\beta_{bc}|=1$ ($s=0.8$), and $r=0.91$, in good agreement with the solutions of Eq.\ref{eq:Sdef}. The device exhibits a direct gain of $18~\text{dB}$ in the forward direction (with a $7~\text{MHz}$ bandwidth), unity transmission in the reverse direction and is impedance matched to the $50~\Omega$ environment, with return loss exceeding  $10~\text{dB}$ over a $4~\text{MHz}$ bandwidth. At this gain, the input power at the $1~\text{dB}$ compression point is $-125~\text{dBm}$ (See Appendix C and Fig.~\ref{figCompression}).

\section{Quadrature sensitivity and noise performance}

Importantly, note that $\mathbf{S_{ac}}$ is only a subset of the complete scattering matrix $\mathbf{S}$, the latter being necessary to understand the quadrature sensitivity of the gain, as well as the added noise of the amplifier. By defining the quadratures of mode $a$
as $X_a= (a^{\text{S}}+a^{\text{I*}})/\sqrt{2}$, $Y_a= i(a^{\text{I*}}-a^{\text{S}})/\sqrt{2}$, and similarly for modes $b$ and $c$, we compute the simplified scattering matrix $\mathbf{Q_{ac}}$ in the quadrature basis $\left( X_a,Y_a,X_c,Y_c \right)$, still in the resonant and lossless case:
\begin{equation}
\mathbf{Q_{ac}}=
\begin{pmatrix}
\ 0\ \ \ &      0\ \ &        0      & \sqrt{G_X}  \\
\ 0\ \ \ &      0\ \ &    \sqrt{G_Y} &      0      \\
\ 0\ \ \ &      1\ \ &        0      &      0	  \\
\ 1\ \ \ &      0\ \ &        0      &      0      
  	\end{pmatrix},
\label{eq:Smatrix}
\end{equation}
where the forward quadrature gains are given by:

\begin{align}
\begin{split}
\sqrt{G_{X}}&=\left(\frac{2s}{1-r}-1\right),\\
\sqrt{G_{Y}}&=\left(\frac{2s}{1+r}-1\right).
\end{split}
\label{eq:Gains}
\end{align}

\begin{figure}[t]
	\includegraphics[scale=1]{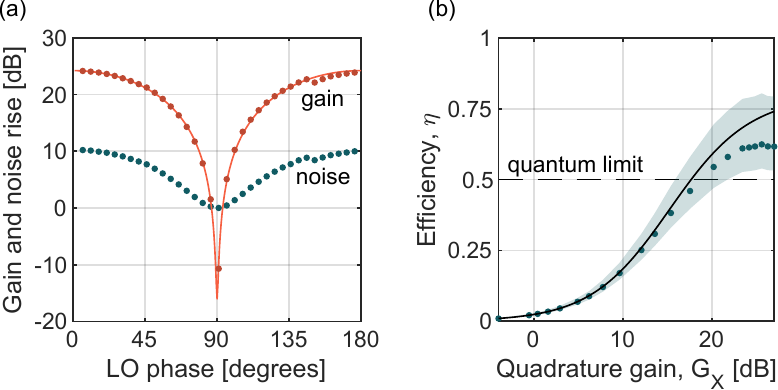}
	\caption{(a) Quadrature gain and noise rise measured for the same pump parameters as in Fig.\ref{fig2}, as a function of the LO phase. We observe a quadrature gain of $24~\text{dB}$ and anti-gain of $-11~\text{dB}$, in good agreement with theoretical predictions from Eq.~\ref{eq:Gains} (line). The measurement noise floor rises much slower than the gain, with up to $14~\text{dB}$ of signal-to-noise improvement. (b) Estimated system efficiency (dots), error bars (shaded area, dominated by the uncertainty of the calibration of $n_{\text{add}}^{\text{chain}}$), and theoretical predictions from Eq.~\ref{eq:Nadd} (lines), as a function of the quadrature gain.
		\label{fig3}}
\end{figure}

The phase-sensitivity is evident, with $G_X\neq G_Y$. Ideal squeezing, defined as $G_XG_Y=1$ \cite{caves1982quantum}, is achieved in the limit $s=1$, \ie $\abs{\beta_{ab}}=\abs{\beta_{bc}} \to \infty$, corresponding to frequency conversion rates to mode $b$ overwhelming its loss rate. Experimentally, starting from the same pump configuration as in Fig.~\ref{fig2}, we measure the quadrature gain by driving a single quadrature at the frequency $\omega_{bb}^p/2 + \omega_{bc}^p\approx\omega_{c}$ and demodulating the output using a mixer and a local oscillator (LO) at the frequency $\omega_{bb}^p/2 - \omega_{ab}^p\approx\omega_{a}$. The results are shown in Fig.~\ref{fig3}(a) and we observe a quadrature gain $G_X\approx24~\text{dB}$ and anti-gain $G_Y\approx-11~\text{dB}$, in good agreement with the theory.

To understand the noise properties of the amplifier, one needs to consider the full scattering matrix $\mathbf{S}$ and the noise at each port. Indeed, the added noise of a parametric amplifier always originates from the noise in additional degrees of freedom. In this work, these degrees of freedom are the signal and idler in mode $b$ (or equivalently the quadratures $X_b$ and $Y_b$). Intuitively, for finite values of $\abs{\beta_{ab}}$ the frequency conversion rates to mode $b$ do not fully overwhelm its loss rate, and noise entering mode $b$ gets amplified alongside signals entering mode $c$, contributing to added noise and limiting squeezing. Formally, one can express the added noise of the amplifier in photon units, $n_{\text{add}}^{\text{FPJA}}$, in the resonant and lossless case, as

\begin{equation}
n_{\text{add}}^{\text{FPJA}}=\frac{1}{8|\beta_{ab}|^{2}}\left(1+G_{X}^{-\frac{1}{2}}\right)^{2}.
\label{eq:Nadd}
\end{equation}

Noiseless amplification is achieved, together with pure squeezing, in the limit of high conversion rate to mode $b$, $\abs{\beta_{ab}} \to \infty$. Experimentally, the system-added noise referred to the input of the amplifier is $n_{\text{add}}=n_{\text{add}}^{\text{FPJA}}+n_{\text{add}}^{\text{chain}}/G_X$, were $n_{\text{add}}^{\text{chain}}=19.8_{-3.3}^{+3.2}$ is the noise added by the measurement chain following the amplifier, calibrated in a separate experiment using the shot noise emitted by a metallic tunnel junction \cite{lecocq2017nonreciprocal}. By comparing the measured noise rise and quadrature gain we extract the improvement in signal-to-noise ratio and convert it into an equivalent system-added noise $n_{\text{add}}$ and measurement efficiency $\eta^{-1}=1+2n_{\text{add}}$. In Fig.~\ref{fig3}(b) we show the measured efficiency $\eta$ as a function of the quadrature gain (dots). At low gain, the efficiency is limited by the following stages of amplification. At high gain the efficiency plateaus, revealing the intrinsic efficiency of the amplifier, $\eta= 0.62_{-0.09}^{+0.17}$, exceeding the quantum limit for phase-preserving amplification. Data are in reasonable agreement with the prediction from Eq. \ref{eq:Nadd} (solid line), within the error bars dominated by the uncertainty of the calibration of $n_{\text{add}}^{\text{chain}}$.

\section{Discussion and conclusion}

Looking forward, lower added noise and larger squeezing can be obtained by reducing the dissipation of mode $b$, \ie $\kappa_{b} \to 0$, or by increasing the coupling rates $g_{ab}$ and $g_{bc}$. The first option is technically limited by the dielectric loss in mode $b$, here estimated to be $1-2~\text{MHz}$. The second option is fundamentally limited by the stability of the amplifier (See Appendix B). To prevent free-oscillation of the amplifier, the amplification rate is limited to $g_{bb} < \kappa_a+\kappa_b+\kappa_c$, or equivalently $4\abs{\beta_{ab}}^2 = 4\abs{\beta_{bc}}^2 < (\kappa_a+\kappa_c)/\kappa_b$ . This puts bounds on the amount of squeezing, added noise, and efficiency of the amplifier:
\begin{equation}
\sqrt{G_Y} > \frac{\kappa_b}{\kappa_a+\kappa_b+\kappa_c},
\label{eq:Gymin}
\end{equation}

\begin{equation}
n_{\text{add}}^{\text{FPJA}} > \frac{\kappa_b}{2\left(\kappa_a+\kappa_c\right)},
\label{eq:naddmin}
\end{equation}

\begin{equation}
\eta < \frac{\kappa_a+\kappa_c}{\kappa_a+\kappa_b+\kappa_c}.
\end{equation}

One can see that the performance of the amplifier only depends on the ratio between the dissipation rate of mode $b$ and the dissipation rates of mode $a$ and $c$. Straightforward improvement of the device design would increase $\kappa_a$ and $\kappa_c$ by a factor of two and decrease $\kappa_b$ by a factor of five, leading to a tenfold decrease in added noise and an efficiency above $0.95$. Additionally larger bandwidth could be obtained by operating at a lower gain, inserting the device as a pre-amplifier in front of a typical wideband and reciprocal parametric amplifier.

In conclusion, we have proposed and experimentally demonstrated, for the first time, a microwave amplifier that is both nonreciprocal and phase-sensitive, demonstrating an efficiency beyond the quantum limit. As a consequence, this amplifier no longer requires a magnetic circulator at its input and can be directly integrated on-chip with a superconducting quantum device. At a quadrature gain of $G_{X}=20~\text{dB}$, this first generation amplifier has a bandwidth of $10~\text{MHz}$ and an input power of $-120~\text{dBm}$ at the $1~\text{dB}$ compression point, already meeting the requirements for a number of applications. For example, this is sufficient for the measurement of a transmon qubit using a coherent state of more than 100 photons in a $1~\text{MHz}$ wide readout cavity \cite{Hatridge2013,Kindel2016}. In addition, this amplifier could enhance quantum state tomography of itinerant microwave fields \cite{Kindel2016}, directly validate continuous variable entanglement \cite{flurin2012entanglement}, and enable position measurements beyond the quantum limit \cite{lecocq2015bae}.

\section*{Acknowledgment}
This work was supported by the NIST Quantum Information Program. This article is a contribution of the U.S. government, not subject to copyright.

\section*{Appendix A: Device description}

\begin{figure}
	\includegraphics[scale=1]{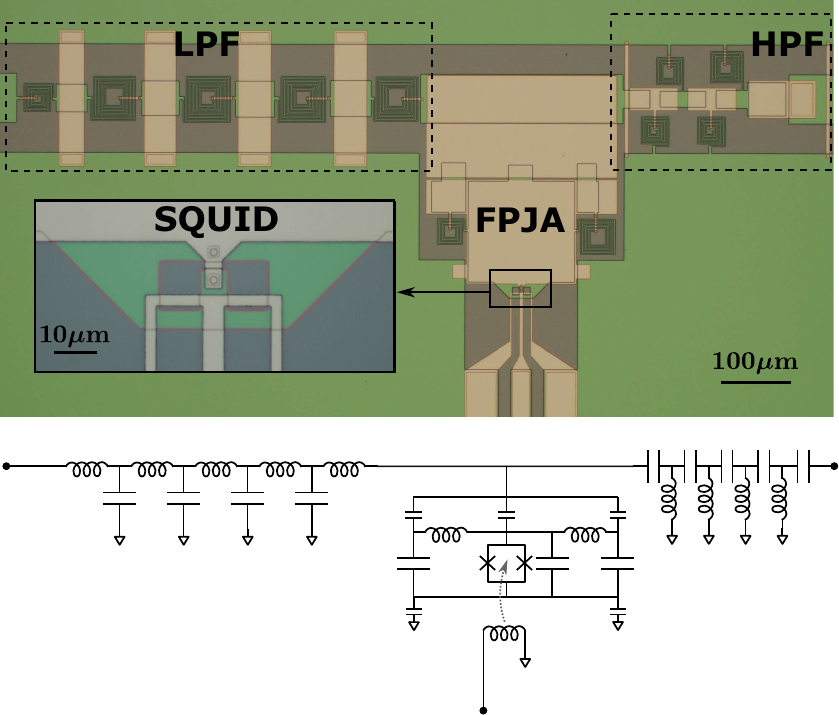}
	\caption{Device picture and layout. (a) Optical micrograph of the device. (b) Circuit equivalent of the device. The center element is a FPJA \cite{lecocq2017nonreciprocal} and consists of a SQUID shunted by a set of inductors and capacitors. The FPJA is capacitively coupled to ground and to an on-chip frequency diplexer built by a 9-pole Low-Pass Filter (LPF), on the left, and a 9-pole High-Pass Filter (HPF), on the right, with a cut-off frequency around $7.6~\text{GHz}$.
		\label{figDevice}}
\end{figure}

An optical micrograph of the device and the corresponding circuit schematic are shown in Fig.~\ref{figDevice}. The device is fabricated using a $\mathrm{Nb/Al-AlO_x/Nb}$ trilayer process with amorphous silicon (a-Si) interlayer dielectric. This process is similar to the one described in earlier work \cite{lecocq2017nonreciprocal}, with one difference: the top $300~\text{nm}$ $\mathrm{Nb}$ wiring layer is replaced by a $\mathrm{Nb/Al/Nb}$ trilayer, with respective thicknesses of $8~\text{nm}$, $8~\text{nm}$ and $300~\text{nm}$. The thin $\mathrm{Al}$ layer acts as an etch stop for the plasma etch of the $\mathrm{Nb}$ wiring layer, while the thin $\mathrm{Nb}$ layer prevents intermixing between the (a-Si) and $\mathrm{Al}$ layers. The patterning of the wiring layer is finalized by ion milling to mechanically remove the exposed thin layers of $\mathrm{Nb}$ and $\mathrm{Al}$.

\begin{figure}
	\includegraphics[scale=1]{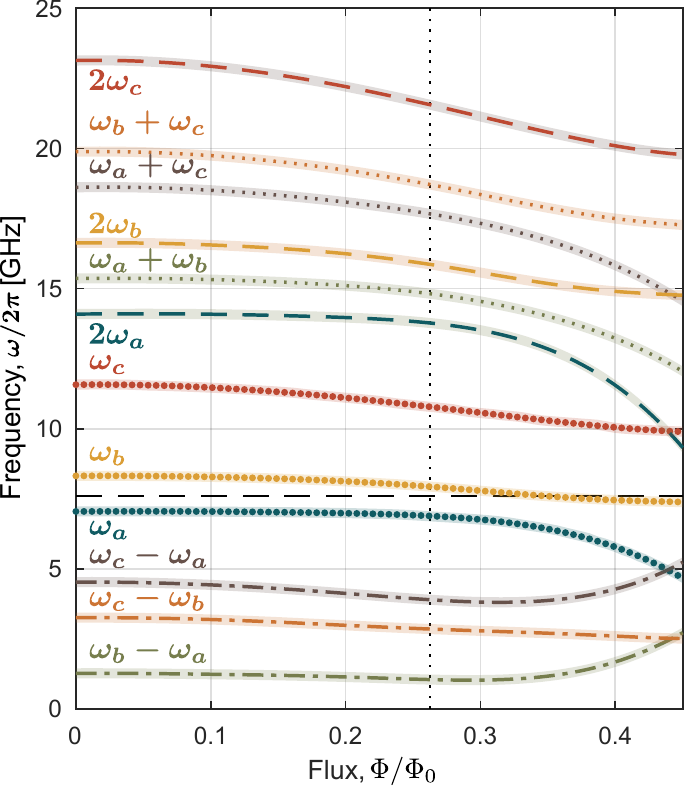}
	\caption{Process spectroscopy. (a) Measured mode frequencies $\omega_{j}$, and predicted first-order modulation frequencies $\abs{\omega_{j}\pm\omega_{k}}$ (${j},{k} \in \{{a,b,c}\}$) as a function of flux. The shaded areas represent a bandwidth of $180~\text{MHz}\approx 3\times(\kappa_a+\kappa_c)/2$, necessary to ensure a good rotating wave approximation. The cut-off of the diplexer at $7.6~\text{GHz}$ is shown as the horizontal dashed line. The circuit is biased at $\Phi/\Phi_0\approx0.26$, shown as a vertical dotted line.
		\label{figModCurve}}
\end{figure}

The central element is the FPJA, a lumped element circuit with three resonance frequencies that all depend on the inductance of a single SQUID. The resonance frequencies as a function of the flux through the SQUID are shown in Fig.~\ref{figModCurve}. Resonators $a$ and $c$ are routed to different physical ports using low-pass and high-pass filters, forming an on-chip diplexer with a cut-off frequency around $7.6~\text{GHz}$.

The mode frequencies were carefully chosen so that all possible first-order modulation frequencies $\abs{\omega_{j}\pm\omega_{k}}$ (${j},{k} \in \{{a,b,c}\}$) were well separated. This ensures  that a good rotating wave approximation can be made for each parametric process. For example, a pump of frequency $\omega_{c}-\omega_{a}$ \textit{only} couples via frequency conversion the signals $a^\text{S}\leftrightarrow c^\text{S}$ and the signals $a^\text{I*}\leftrightarrow c^\text{I*}$, but no other signals.

\section*{Appendix B: Detailed theory}

In this section we described the calculations of the scattering parameters, output noise and stability of the amplifier, following concepts and style of \cite{lecocq2017nonreciprocal}.

\subsubsection{Scattering parameters in the mode basis}

The device can be described as three modes with natural frequencies $\omega_{a}$, $\omega_{b}$ and $\omega_{c}$, coupled to each other via a tunable inductor (the SQUID). We apply four pumps to modulate the inductance at the frequencies $\omega_{ab}^p \approx \abs{\omega_{b}-\omega_{a}}$, $\omega_{bc}^p \approx \abs{\omega_{c}-\omega_{b}}$, $\omega_{ac}^p \approx \abs{\omega_{c}-\omega_{a}}$ and $\omega_{bb}^p \approx \abs{\omega_{b}+\omega_{b}}$, effectively coupling signals at six different frequencies, $\omega_{a}^\text{S}$, $\omega_{b}^\text{S}$, $\omega_{c}^\text{S}$, $\omega_{a}^\text{I}$, $\omega_{b}^\text{I}$ and $\omega_{c}^\text{I}$.

We define the vector of intra-cavity field amplitudes $\mathbf{A}$, input field amplitudes $\mathbf{A}_{\text{in}}$, output field amplitudes $\mathbf{A}_{\text{out}}$, diagonal matrices for the total loss rates $\mathbf{K}$, external couplings $\mathbf{K}^{\text{ext}}$ and coupling efficiencies $\mathbf{H}$: 

\begin{align}
\begin{split}
\mathbf{A}&=(a^\text{S},b^\text{S},c^\text{S},a^\text{I*},b^\text{I*},c^\text{I*})^\top,\\
\mathbf{A}_{\text{in}}&=(a_\text{in}^\text{S},b_\text{in}^\text{S},c_\text{in}^\text{S},a_\text{in}^\text{I*},b_\text{in}^\text{I*},c_\text{in}^\text{I*})^\top,\\
\mathbf{A}_{\text{out}}&=(a_\text{out}^\text{S},b_\text{out}^\text{S},c_\text{out}^\text{S},a_\text{out}^\text{I*},b_\text{out}^\text{I*},c_\text{out}^\text{I*})^\top,\\
\mathbf{K}&=\text{diag}(\sqrt{\kappa_a},\sqrt{\kappa_b},\sqrt{\kappa_c},\sqrt{\kappa_a},\sqrt{\kappa_b},\sqrt{\kappa_c}),\\
\mathbf{K}^{\text{ext}}&=\text{diag}(\sqrt{\kappa_a^{\text{ext}}},\sqrt{\kappa_b^{\text{ext}}},\sqrt{\kappa_c^{\text{ext}}},\sqrt{\kappa_a^{\text{ext}}},\sqrt{\kappa_b^{\text{ext}}},\sqrt{\kappa_c^{\text{ext}}}),\\
\mathbf{H}&=\text{diag}(\sqrt{\eta_a},\sqrt{\eta_b},\sqrt{\eta_c},\sqrt{\eta_a},\sqrt{\eta_b},\sqrt{\eta_c}),\\
\end{split}
\label{eq:ModeVectorDef}
\end{align}

where  $\eta_{j}=\kappa_{j}^{\text{ext}}/\kappa_{j}$. Note that, as in \cite{lecocq2017nonreciprocal}, we study here the dynamics of the expectation values of the field operators, defining $a^\text{S} \equiv \left< \hat{a}^\text{S}\right>$ and $a^{\text{I*}} \equiv \left< \hat{a}^{\text{I}\dagger}\right>$ (and similarly for $b^\text{S}$, $b^{\text{I*}}$,$c^\text{S}$ and $c^{\text{I*}}$).

The Equations of Motion (EoM) can be expressed as $\mathbf{K}\mathbf{M}\mathbf{K}\mathbf{A} = i\mathbf{K}^{\text{ext}}\mathbf{A}_{\text{in}}$, where  $\mathbf{M}$ is the mode-coupling matrix:

\begin{equation}
\mathbf{M}=
\left(\begin{array}{ccc|ccc}
\ \Delta_{a}^S   & \   \  \beta_{ab}  &\ \ \beta_{ac}   &     0           &        0        &      0     \\
\ \beta_{ab}^{*} & \ \Delta_{b}^S   &\ \  \beta_{bc}   &     0           &    \beta_{bb}   &      0     \\
\ \beta_{ac}^*   & \ \beta_{bc}^{*} &\ \  \Delta_{c}^S &     0           &        0        &      0     \\ \hline
\ 0         &  \  \     0        &\   \     0       & -{\Delta_a^I}^* & -\beta_{ab}^*   & -\beta_{ac}^* \\
\ 0         & \ -\beta_{bb}^*  &\   \     0       & -\beta_{ab}     & -{\Delta_b^I}^* & -\beta_{bc}^* \\
\ 0         &  \  \     0        &  \  \    0       & -\beta_{ac}     & -\beta_{bc}     & -{\Delta_c^I}^*
\end{array}\right).
\label{eq:BTmatrix}
\end{equation}

\begin{figure}
	\includegraphics{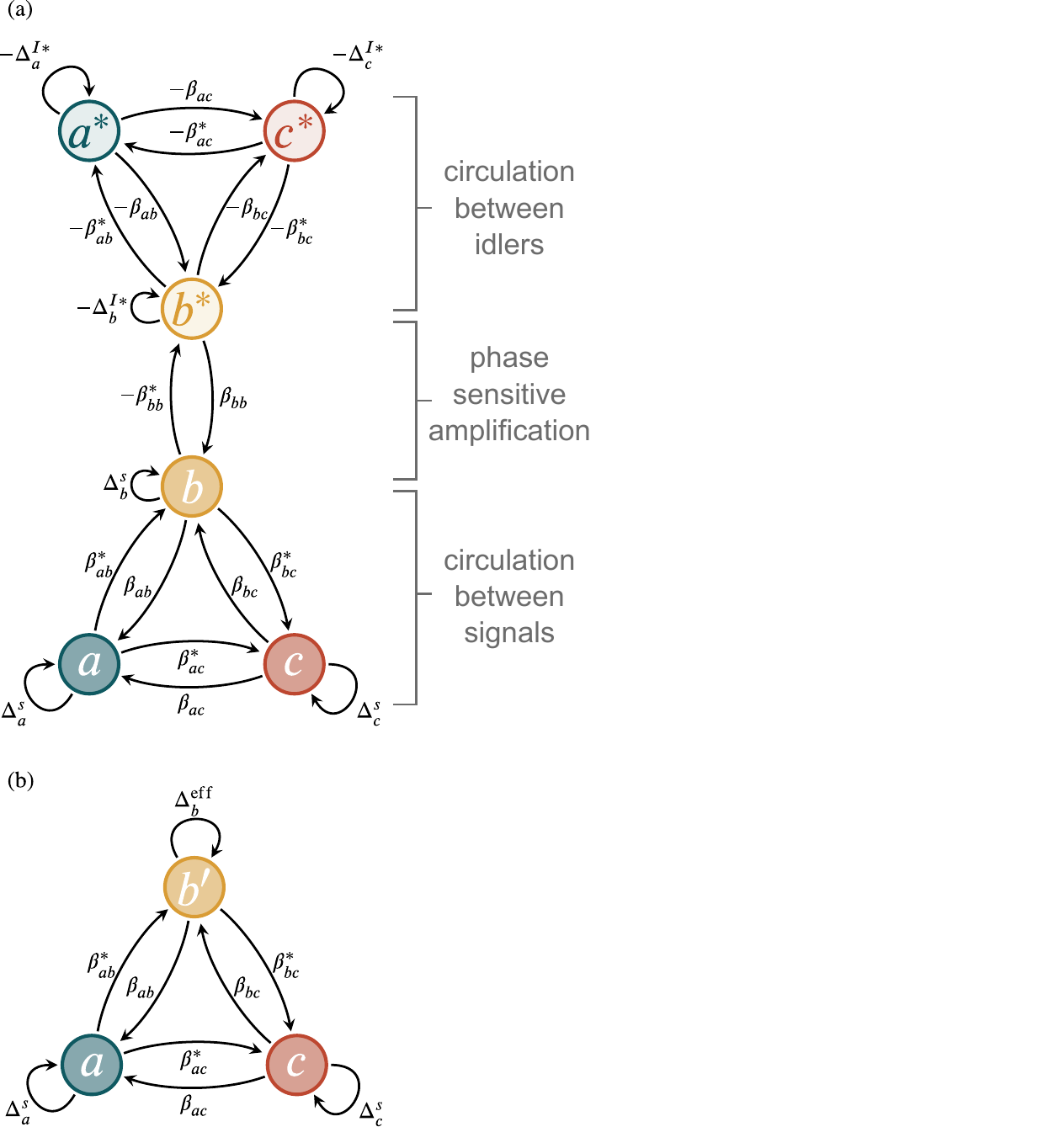} \caption{Mode coupling graphs. (a) The full graph for the phase-sensitive directional amplifier. (b) Reduced graph evaluated at zero pump detuning. This
		graph is equivalent to that in (a), with the upper triangle of the
		original graph included as an effective self-loop of mode $b$. }
	\label{fig:graph} 
\end{figure}

The diagonal elements are the normalized complex detuning terms $\Delta_j^{S,I}=(\omega_{j}^\text{S,I}-\omega_{j})/\kappa_j+i/2$ and the off-diagonal elements are the normalized coupling terms 
$\beta_{jk}=g_{jk}/(2\sqrt{\kappa_j\kappa_k})$ between modes ${j}$ and ${k}$, with ${j},{k} \in \{{a,b,c}\}$. Each $g_{jk}=\abs{g_{jk}}e^{i\phi_{jk}}$ is the parametrically induced coupling rate, proportional to the amplitudes of the pump with frequency $\omega_{jk}^p$ and phase $\phi_{jk}$.  

We emphasize the block structure of $\mathbf{M}$, which reflects the coupling networks in Fig.~\ref{fig:graph}a and frequency space diagram in Fig.~\ref{figTuning}(c,d): the diagonal blocks correspond to circulation between signals or between idlers, and the anti-diagonal blocks correspond to amplification between signal and idler of mode $b$.

Solving these EoM yields the scattering matrix $\mathbf{S}$, defined as $\mathbf{A}_{\text{out}}=\mathbf{S}\mathbf{A}_{\text{in}}$:
\begin{equation}
\mathbf{S}= i\mathbf{H}\mathbf{M}^{-1}\mathbf{H}-\mathds{1}.
\label{eq:Sdef}
\end{equation}

We now derive analytical expressions for the scattering parameters when all four pumps are on resonance (that is $\omega_{ab}^{p}=\omega_{b}-\omega_{a}$, $\omega_{bc}^{p}=\omega_{c}-\omega_{b}$, $\omega_{ac}^{p}=\omega_{c}-\omega_{a}$, and $\omega_{bb}^{p}=2\omega_{b}$). The resonance conditions mean $-\Delta_{j}^{I*}=\Delta_{j}^{S}$, so the matrix $\mathbf{M}$ from Eq.~\ref{eq:BTmatrix} becomes much simpler. To simplify notations we will omit in the following the signal and idler superscript for the detuning terms $\Delta_{j}^\text{S,I}\rightarrow\Delta_{j}$.

We are most interested in the $4\times4$ scattering matrix between
the signal and idler modes in the ``input'' mode $c$ and ``output'' mode $a$. We can write
some of these scattering elements compactly as
\begin{align}
S_{a^\text{S}a^\text{S}} & =\frac{i\eta_{a}C}{|\mathbf{M}|}\left(\Delta_{b}^{\text{eff}}\Delta_{c}-|\beta_{bc}|^{2}\right)-1,\label{eq:Saa}\\
S_{c^\text{S}c^\text{S}} & =\frac{i\eta_{c}C}{|\mathbf{M}|}\left(\Delta_{b}^{\text{eff}}\Delta_{a}-|\beta_{ab}|^{2}\right)-1,\label{eq:Scc}\\
S_{a^\text{I*}a^\text{S}} & =\frac{i\eta_{a}}{|\mathbf{M}|}\beta_{bb}^{*}\left(\beta_{bc}^{2}\beta_{ac}^{*2}-\beta_{ab}^{*2}\Delta_{c}^{2}\right),\label{eq:SaIaS}\\
S_{c^\text{I*}c^\text{S}} & =\frac{i\eta_{c}}{|\mathbf{M}|}\beta_{bb}^{*}\left(\beta_{ac}^{2}\beta_{ab}^{*2}-\beta_{bc}^{2}\Delta_{a}^{2}\right),\label{eq:ScIcS}\\
S_{a^\text{S}c^\text{S}} & =\frac{i\sqrt{\eta_{a}\eta_{c}}C}{|\mathbf{M}|}\left(\beta_{ab}\beta_{bc}-\beta_{ac}\Delta_{b}^{\text{eff}}\right),\label{eq:Sac}\\
S_{c^\text{S}a^\text{S}} & =\frac{i\sqrt{\eta_{a}\eta_{c}}C}{|\mathbf{M}|}\left(\beta_{ab}^{*}\beta_{bc}^{*}-\beta_{ac}^{*}\Delta_{b}^{\text{eff}}\right),\label{eq:Sca}\\
S_{a^\text{I*}c^\text{S}}& =\frac{i\sqrt{\eta_{a}\eta_{c}}}{|\mathbf{M}|}\beta_{bb}\left(\beta_{ab}\Delta_{c}-\beta_{ac}\beta_{bc}^{*}\right)\left(\beta_{bc}^{*}\Delta_{a}+\beta_{ab}\beta_{ac}^{*}\right). \label{eq:SaIcS}
\end{align}
where
\begin{multline} 
C = \Delta_{a}\Delta_{b}\Delta_{c}-|\beta_{bc}|^{2}\Delta_{a}-|\beta_{ac}|^{2}\Delta_{b}-|\beta_{ab}|^{2}\Delta_{c}\\
-\beta_{ab}\beta_{bc}\beta_{ac}^{*}-\beta_{ab}^{*}\beta_{bc}^{*}\beta_{ac}
\end{multline}
is the determinant of the $a^\text{I*}\rightarrow b^\text{I*}\rightarrow c^\text{I*}$
loop, and $\Delta_{b}^{\text{eff}}$ is an effective detuning for the field amplitude $b^\text{S}$  that includes the dynamics of all the idler fields (\ie reducing out the upper triangle in Fig.~\ref{fig:graph}a to obtain Fig.~\ref{fig:graph}b): 
\begin{equation}
\Delta_{b}^{\text{eff}}=\Delta_{b}+\frac{|\beta_{bb}|^{2}}{C}\left(\Delta_{a}\Delta_{c}-|\beta_{ac}|^{2}\right).\label{eq:EffectiveDelta}
\end{equation}

$S_{a^\text{S}a^\text{S}}$ and $S_{c^\text{S}c^\text{S}}$ are the reflection coefficient off of mode $a$ and $c$ respectively. $S_{a^\text{S}c^\text{S}}$ and $S_{c^\text{S}a^\text{S}}$ are respectively the forward and reverse gain and $S_{a^\text{I*}c^\text{S}}$ is the forward idler gain.

Imposing the condition $S_{a^\text{I*}a^\text{S}}=S_{c^\text{I*}c^\text{S}}=0$ on resonance
(that is, no transmission from signal to idler of the same resonance),
we find $|\beta_{ab}|=|\beta_{bc}|$, $|\beta_{ac}|=1/2$ and $\phi_\text{loop}=\phi_{ab}+\phi_{bc}-\phi_{ac}=\pm\pi/2$.
In the following, we choose $\phi_\text{loop}=\pi/2$ to fix the direction of
circulation from $c^\text{S}\rightarrow b^\text{S}\rightarrow a^\text{S}$. Substituting these conditions into the scattering parameters above, and evaluating on resonance ($\Delta_j=i/2$), we can calculate the scattering matrix for the reduced basis $(a^\text{S},c^\text{S},a^\text{I*},c^\text{I*})$ as 
\begin{equation}
\mathbf{S_{ac}}=\left(\begin{array}{cccc}
\eta_{a}-1 & -i\sqrt{G_\text{S}} & 0 & -e^{i\phi_{bb}}\sqrt{G_\text{I}}\\
i\sqrt{\eta_{a}\eta_{c}} & \eta_{c}-1 & 0 & 0\\
0 & -e^{-i\phi_{bb}}\sqrt{G_\text{I}} & \eta_{a}-1 & i\sqrt{G_\text{S}}\\
0 & 0 & -i\sqrt{\eta_{a}\eta_{c}} & \eta_{c}-1
\end{array}\right),\label{eq:app_Sac}
\end{equation}
where
\begin{align}
\sqrt{G_\text{S}} & =\sqrt{\eta_{a}\eta_{c}}\frac{2s+r^{2}-1}{1-r^{2}},\\
\sqrt{G_\text{I}} & =\sqrt{\eta_{a}\eta_{c}}\frac{2rs}{1-r^{2}},
\end{align}
and 
\begin{align}
s & =\frac{4|\beta_{ab}|^{2}}{1+4|\beta_{ab}|^{2}},\\
r & =\frac{2|\beta_{bb}|}{1+4|\beta_{ab}|^{2}},
\end{align}
with $0\leq r,s\leq1$.

Note the differences between Eq.~\ref{eq:app_Sac} and the corresponding equation Eq.~3 in the main text. In the main text we simplified $\mathbf{S_{ac}}$ by writing only the magnitude of the matrix elements, and used a reduced basis that excludes the idlers.

\subsubsection{Scattering parameters in the quadrature basis}

We rotate from the mode basis $\mathbf{A}$, $\mathbf{A}_{\text{in}}$, $\mathbf{A}_{\text{out}}$ to the quadrature basis $\mathbf{X}=\mathbf{U}\mathbf{A}$, $\mathbf{X}^{\text{in}}=\mathbf{U}\mathbf{A}_{\text{in}}$, $\mathbf{X}^{\text{out}}=\mathbf{U}\mathbf{A}_{\text{out}}$ where $\mathbf{U}$ is the unitary matrix 

\begin{equation}
\mathbf{U}=\frac{1}{\sqrt{2}}\left(\begin{array}{cccccc}
1 & 0 & 0 & \ 1 &\ \  0 &\ \  0\\
-i & 0 & 0 & \ i &\ \  0 &\ \  0\\
0 & 1 & 0 & \  0 & \ \ 1 &\ \  0\\
0 & -i & 0 & \  0 &\ \  i & \ \ 0\\
0 & 0 & 1 &  \ 0 &\ \  0 &\ \  1\\
0 & 0 & -i &  \  0 &\ \  0 &\ \  i
\end{array}\right)
\end{equation}
and 
\begin{align}
\begin{split}
\mathbf{X}&=(X_a,X_b,X_c,Y_a,Y_b,Y_c)^\top,\\
\mathbf{X}_\text{in}&=(X_{a,\text{in}},X_{b,\text{in}},X_{c,\text{in}},Y_{a,\text{in}},Y_{b,\text{in}},Y_{c,\text{in}})^\top,\\
\mathbf{X}_\text{out}&=(X_{a,\text{out}},X_{b,\text{out}},X_{c,\text{out}},Y_{a,\text{out}},Y_{b,\text{out}},Y_{c,\text{out}})^\top.\\
\end{split}
\label{eq:QuadVectorDef}
\end{align}

The scattering matrix in the quadrature basis, $\mathbf{Q}$, defined as $\mathbf{X}_{\text{out}}=\mathbf{Q}\mathbf{X}_{\text{in}}$, is $\mathbf{Q}=\mathbf{U}\mathbf{S}\mathbf{U}^{-1}$. To line up the quadrature definitions with the phase-sensitive gain,
we choose $\phi_{bb}=-\pi/2$. The scattering matrix for the reduced quadrature basis
$(X_{a},Y_{a},X_{c},Y_{c})$ is then:

\begin{equation}
\mathbf{Q_{ac}}=\left(\begin{array}{cccc}
\eta_{a}-1 & 0 & 0 & \sqrt{G_{X}}\\
0 & \eta_{a}-1 & -\sqrt{G_{Y}} & 0\\
0 & -\sqrt{\eta_{a}\eta_{c}} & \eta_{c}-1 & 0\\
\sqrt{\eta_{a}\eta_{c}} & 0 & 0 & \eta_{c}-1
\end{array}\right),\label{eq:Squad}
\end{equation}
where 
\begin{equation}
\sqrt{G_{X}}=\sqrt{G_\text{S}}+\sqrt{G_\text{I}}=\sqrt{\eta_{a}\eta_{c}}\left(\frac{2s}{1-r}-1\right),
\label{eq:GainsAppendix}
\end{equation}
and 
\begin{equation}
\sqrt{G_{Y}}=\sqrt{G_\text{S}}-\sqrt{G_\text{I}}=\sqrt{\eta_{a}\eta_{c}}\left(\frac{2s}{1+r}-1\right).
\end{equation}

Note that in the main text we simplified $\mathbf{Q_{ac}}$ (Eq.6) by writing only the magnitude of the matrix elements of Eq.~\ref{eq:Squad}.

The product of the two quadrature gains is 
\begin{equation}
\sqrt{G_{X}G_{Y}}=\eta_{a}\eta_{c}\left(1-4s\frac{1-s}{1-r^{2}}\right).
\end{equation}

The two free parameters, $r$ and $s$, are determined by the two remaining free coupling rates $|\beta_{bb}|$ and $|\beta_{ab}|=|\beta_{bc}|$. Ideal squeezing, corresponding to $G_XG_Y=1$ \cite{caves1982quantum}, is achieved in the limit $\eta_{a}=\eta_{c}=1$ and $s=1$, \ie $\abs{\beta_{ab}}=\abs{\beta_{bc}} \to \infty$.

\subsubsection{Amplifier output noise}

To compute the output noise of the amplifier, we apply the scattering matrix to general input states that includes vacuum fluctuations and thermal contributions at every port. To that end we redefine here the vectors of field and quadrature amplitudes to be vectors of \textit{operators}, ensuring the proper commutation relations. The output covariance matrix in terms of the field quadratures is 
\begin{equation}
\langle\mathbf{X}_\text{out}^\dagger\mathbf{X}_\text{out}^\top\rangle=\mathbf{U}^{*}\mathbf{S}^{*}\langle\mathbf{A}_{\text{in}}^{\dagger}\mathbf{A}_{\text{in}}^{\top}\rangle\mathbf{S}\mathbf{U},
\end{equation}
where the input covariance matrix $\langle\mathbf{A}_{\text{in}}^{\dagger}\mathbf{A}_{\text{in}}{}^{\top}\rangle$
is diagonal with elements determined by the input thermal state occupancies. While general analytical solutions
exist, it is more useful to look at relevant limits. Evaluating the
quadrature output noise on resonance for lossless modes with vacuum-state
inputs, we find
\begin{align}
\langle X_{a,\text{out}}^{\dagger}X_{a,\text{out}}\rangle & =G_{X}\left[\frac{1}{2}+n_{\text{add}}^{\text{FPJA}}\right],
\end{align}
where the added noise $n_{\text{add}}^{\text{FPJA}}$ can be written in terms of
$G_{X}$ and $|\beta_{ab}|$ as 
\begin{equation}
n_{\text{add}}^{\text{FPJA}}=\frac{1}{8|\beta_{ab}|^{2}}\left(1+G_{X}^{-\frac{1}{2}}\right)^{2}.
\end{equation}
For large gain, the added noise is determined by the coupling strength
to the $b$ mode. For the minimal added noise, $|\beta_{ab}|=|\beta_{bc}|$
must be maximized either by taking $\kappa_{b}\rightarrow0$ or $g_{ab},g_{bc}\rightarrow\infty$. However, $g_{ab}$ and $g_{bc}$ are limited by the stability of the amplifier.

As described in Appendix B.4 and B.5 of \cite{lecocq2017nonreciprocal}, a more complete model can be obtained by generalizing
the scattering matrix to a $12\times12$ matrix that includes internal ports for each mode.

\subsubsection{Amplifier stability}

For the Langevin equations of motion to be stable, all the eigenvalues of the Langevin matrix $\mathbf{M}$ need to have a negative real part corresponding to damping. In the ideal resonant case, neglecting internal loss, and under the conditions $|\beta_{ac}|=1/2$ and $|\beta_{ab}|=|\beta_{bc}|$, the characteristic polynomial  $P(\lambda)=|\mathbf{M}(\omega=i\lambda)|$ takes on a rather simple form:

\begin{equation}
P =P_+P_-+b_\phi,~b_\phi = \kappa_a^2\kappa_b^2\kappa_c^2|\beta_{ab}|^2cos^2\phi_\text{loop}
\end{equation}

where the polynomials $P\pm$ are of third order and independent of the loop phase:
\begin{equation}
P\pm = \lambda^3 + b_1^\pm\lambda^3 + b_2^\pm\lambda^2 + b_3^\pm
\end{equation}

with 

\begin{equation}
\begin{aligned}
b_1^\pm &= \frac{\kappa_{b}}{2}(1 \mp 2\beta_{bb})+\frac{\kappa_{a}+\kappa_{c}}{2},\\
b_2^\pm &= \frac{\kappa_{b}(\kappa_{a}+\kappa_{c})}{4}\left(4|\beta_{ab}|^{2}+1\mp 2\beta_{bb}\right)+\frac{\kappa_{a}\kappa_{c}}{2},\\
b_3^\pm &= \frac{\kappa_{a}\kappa_{b}\kappa_{c}}{4}\left(4|\beta_{ab}|^{2}+1\mp 2\beta_{bb}\right).
\end{aligned}
\end{equation}

The Langevin equations of motion are stable if all these coefficients are positive \cite{gantmacher2005applications}. We notice that $b_{1,2,3}^->0$ and $b_{\phi}=0$ for $\phi_\text{loop}=\pm\pi/2$. Solving $b_{1,2,3}^+>0$ gives the following conditions:

\begin{equation}
\begin{aligned}
\beta_{bb} & <\frac{1}{2}+ 2|\beta_{ab}|^{2},\\
\beta_{bb} & <\frac{1}{2}+\frac{\kappa_{a}+\kappa_{c}}{2\kappa_{b}},\\	
\beta_{bb} & <\frac{1}{2}+2|\beta_{ab}|^{2}+\frac{\kappa_{a}\kappa_{c}}{\kappa_{b}\left(\kappa_{a}+\kappa_{c}\right)}.
\end{aligned}
\end{equation}

The first condition corresponds to the pole $r=1$ in the gain equations (Eq.~\ref{eq:GainsAppendix}), therefore the second and last conditions must be less restrictive than the first one. This implies $|\beta_{ab}|^{2}<(\kappa_{a}+\kappa_{c})/4\kappa_{b}$ and therefore $g_{bb}<\kappa_{a}+\kappa_{b}+\kappa_{c}$. The latter condition has a simple physical interpretation: the parametric amplification rate cannot be greater than the total dissipation rate of the three modes.

The upper limit on $|\beta_{ab}|$ sets the limits on the amount of squeezing, added noise and efficiency:

\begin{equation}
\sqrt{G_Y} > \frac{\kappa_b}{\kappa_a+\kappa_b+\kappa_c}
\end{equation}

\begin{equation}
n_{\text{add}}^{\text{FPJA}} > \frac{\kappa_b}{2\left(\kappa_a+\kappa_c\right)}
\end{equation}

\begin{equation}
\eta < \frac{\kappa_a+\kappa_c}{\kappa_a+\kappa_b+\kappa_c}
\end{equation}

\begin{figure}
	\includegraphics{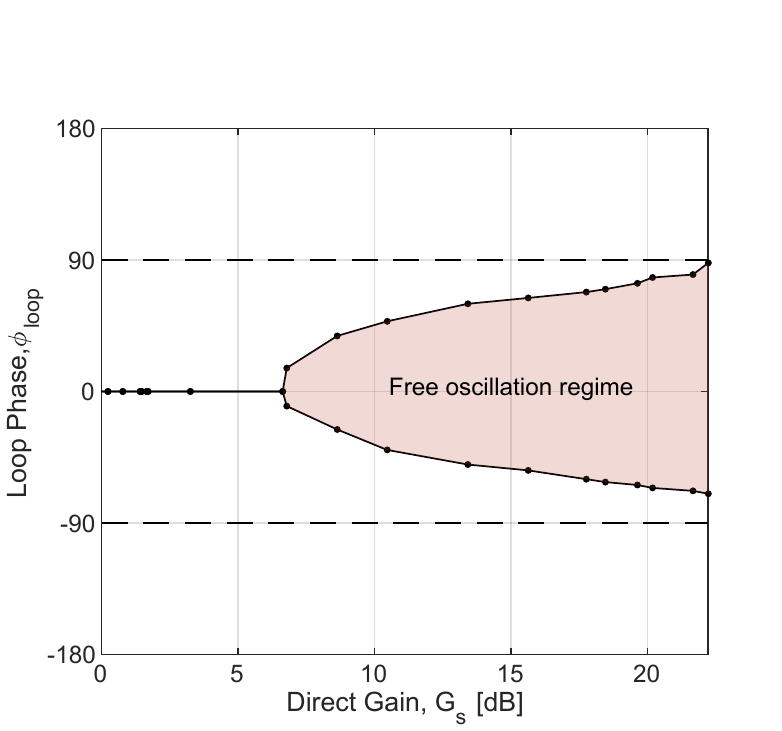} \caption{Measured region of instability as a function of the amplifier direct gain, for the same pump parameters as in Fig.\ref{fig2}. Below a gain of $\approx6~\text{dB}$, the amplifier is stable for all loop phases. At higher gain the region of instability grows and approaches the loop phases $\phi_\text{loop}=\pm\pi/2$ at which the amplifier is directional. \label{figStability}}
\end{figure}

We observe experimentally this instability by measuring the scattering parameters of the amplifier as a function of the loop phase $\phi_\text{loop}$ and the direct gain. The region of instability is shown in Fig.~\ref{figStability}. When the direct gain of the amplifier exceeds $\approx6~\text{dB}$, a region of parametric oscillations appears around $\phi_\text{loop}=0$. This region grows with the gain, approaching at high gain the loop phases $\phi_\text{loop}=\pm\pi/2$ at which the amplifier is directional.
Comparison with theory remains a work in progress and involves finding the roots of the characteristic polynomial $P$ in the more general case of arbitrary pump strengths, frequencies, and phases.  

\section*{Appendix C: Dynamic range}

\begin{figure}
	\includegraphics[scale=1.0]{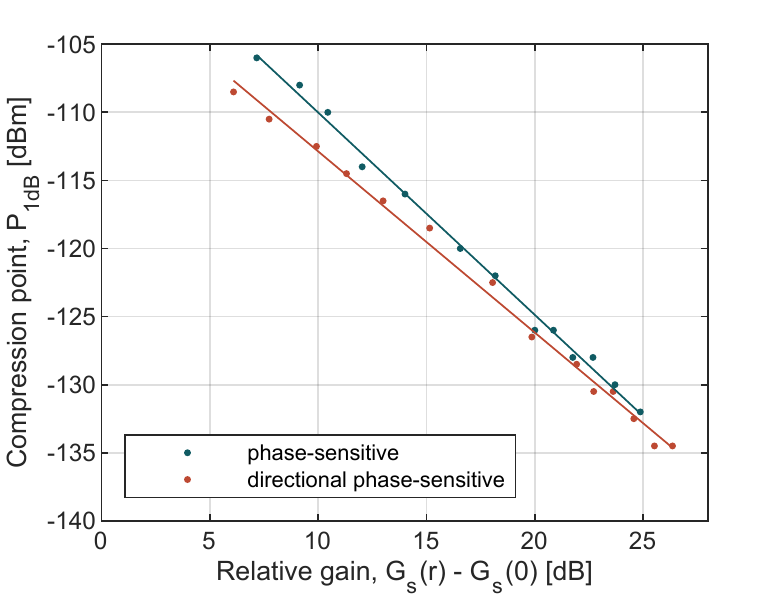}
	\caption{Dynamic range. Input power at the $1~\text{dB}$ compression point, as a function of gain, for two modes of operation of the FPJA. In green, the FPJA is programmed using a single pump at $\omega_{aa}^p = 2\omega_{a}$, leading to phase-sensitive gain around mode $a$. In red, the FPJA is programmed as a directional phase-sensitive amplifier, as described in the main text. The gain is defined as $G_\text{S}(r)-G_\text{S}(0)$, to account for internal losses and frequency conversion losses at finite $s$. The solid green and red lines are linear fits, with respective slope of $-1.5~\text{dBm/dB}$ and $-1.3~\text{dBm/dB}$.
		\label{figCompression}}
\end{figure}

The dynamic range of the amplifier is similar to that of a regular Josephson Parametric Amplifier of similar bandwidth and non-linearity. In Fig.~\ref{figCompression}, we compare the input power at the $1~\text{dB}$ compression point when the FPJA is programmed as a directional phase-sensitive amplifier (in red) or a regular phase-sensitive amplifier (in green). The latter is achieved by using a single pump at $\omega_{aa}^p = 2\omega_{a}$, leading to phase-sensitive gain around mode $a$. The $1~\text{dB}$ compression point is shown as a function of $G_s(r)-G_s(0)$, \ie of the direct gain relative to the scattering amplitude when the amplifier's gain pump is off (pump at $\omega_{aa}^p$ or $\omega_{bb}^p$). This accounts for internal losses and frequency conversion losses for $s<1$. The solid green and red lines are linear fits, with respective slope of $-1.5~\text{dBm/dB}$ and $-1.3~\text{dBm/dB}$.

\section*{Appendix D: Amplifier tuning}

\begin{figure*}[h]
	\includegraphics[scale=1.0]{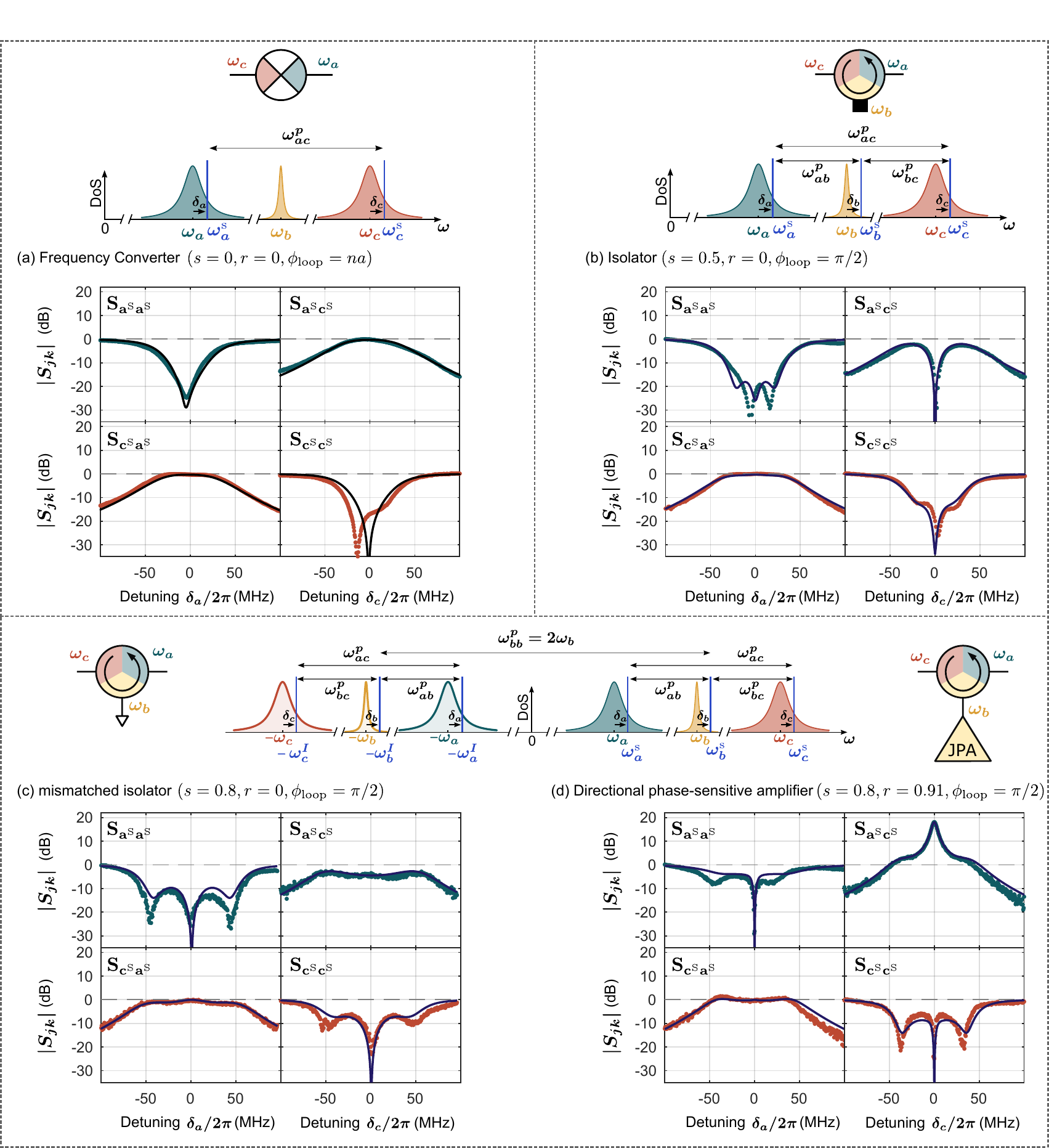}
	\caption{Device tuning. The directional phase-sensitive amplifier is tuned in four steps. (a) Frequency converter. A pump at $\omega_{ac}^p = \abs{\omega_{c}-\omega_{a}}$ enables near ideal frequency conversion between modes $a$ and $c$. (b) Isolator. Two additional pumps at $\omega_{ab}^p = \abs{\omega_{b}-\omega_{a}}$ and $\omega_{bc}^p = \abs{\omega_{c}-\omega_{b}}$, couple the modes $a$ and $c$ to mode $b$ at a rate matching its dissipation. The device behaves as an isolator, with the loop phase set to have unity transmission from mode $a$ to mode $c$ and high isolation in the reversed direction. (c) Mismatched isolator. The coupling rates to mode $b$ are increased to overwhelm its dissipation, leading to almost full reflection off of mode $b$. Impedance matching and unity transmission from mode $a$ to mode $c$ are maintained but isolation in the reversed direction is reduced. This will ensure a low added noise when operated as an amplifier.  (d) Directional phase-sensitive amplifier at high gain. A pump at $\omega_{bb}^p = 2\omega_{b}$ induces gain from mode $c$ to mode $a$.
		\label{figTuning}}
\end{figure*}

As discussed in the main text, the device is programmed in four steps. We describe here in more detail these steps and we show the corresponding measured scattering parameters in Fig.~\ref{figTuning}.

\begin{itemize}
	\item First each frequency conversion pump is individually calibrated to produce ideal conversion, $\beta_{jk}=1/2$. In Fig.~\ref{figTuning}(a) we show the example of the scattering parameters for the case of frequency conversion between modes $a$ and $c$. We observe near-ideal transmission and low reflection. 
	\item Second, all three frequency conversion pumps are turned on. As discussed in detail in \cite{lecocq2017nonreciprocal}, this pump configuration creates a parametric circulator, whose circulation direction is set by the sum of the pump phases, \textit{i.e.} loop phase, $\phi_\text{loop}=\phi_{ab}+\phi_{bc}-\phi_{ac}$. In Fig.~\ref{figTuning}(b), we show the scattering parameters between modes $a$ and $c$ for $\phi_\text{loop}=\pi/2$. We observe unity transmission from $a$ to $c$ and isolation up to $30~\text{dB}$ in the reverse direction. Both modes are still impedance matched, with low reflection coefficients.
	\item Third, the strength of the frequency conversion pumps to mode $b$, $\beta_{ab}$ and $\beta_{bc}$, are increased in order to overwhelm the loss of mode $b$, maintaining $|\beta_{ab}|=|\beta_{bc}|$, see Fig.~\ref{figTuning}(c). This effectively yields high reflection off of mode $b$ (not shown) and the scattering parameters resemble the case of the lossless JPA on a circulator in the absence of gain. We still observe unity transmission from  $a$ to $c$, but the device is almost reciprocal again, with only about $3~\text{dB}$ of isolation in the reverse direction (set by the value of $s=0.8$). Importantly, this pump configuration sets the gain-bandwidth product of the amplifier, set by the width of modes $a$ and $c$, and mostly independent of the width of mode $b$ (in the limit $\kappa_b\ll\kappa_a,\kappa_c$)
	\item Finally, the amplification pump is turned on and corresponding scattering are show in Fig.~\ref{figTuning}(d) (same data than Fig.~2, over a wider frequency span). Gain from $c$ to $a$ grows, while unity transmission is preserved from mode $a$ to $c$. Impedance matching is maintained, over a narrower bandwidth.
\end{itemize}

\end{document}